\documentclass[a4paper,twocolumn,superscriptaddress,aps,prd,10pt]{revtex4-1}
\usepackage{graphicx}
\usepackage{color}
\usepackage{array}

\begin{document}
\title{The Work Functions of Au/Mg Decorated Au(100), Mg(001), and AuMg Alloy Surfaces: A Theoretical Study}
\author{Mat\'u\v{s} Dubeck\'{y}}
\email[]{\tt matus.dubecky@upol.cz}
\affiliation{Regional Centre of Advanced Technologies and Materials,
Department of Physical Chemistry, Faculty of Science, Palack\'y University
Olomouc, t\v{r}.~17~listopadu 12, 771 46 Olomouc, Czech Republic}
\affiliation{Institute of Electrical Engineering, Slovak Academy of Sciences, D\'{u}bravsk\'{a} cesta 9, SK-84104 Bratislava, Slovakia}
\author{Franti\v{s}ek Dubeck\'{y}}
\affiliation{Institute of Electrical Engineering, Slovak Academy of Sciences, D\'{u}bravsk\'{a} cesta 9, SK-84104 Bratislava, Slovakia}

\date{\today}

\begin{abstract}
A plane-wave density functional theory is used to predict the work functions of Au/Mg decorated Au(100), Mg(001), and stochiometric AuMg alloy surfaces. We find, that irrespective of the details, all Au/Mg systems containing Mg on the surface reveal the Mg-dominated work functions, i.e. significantly shifted toward the work function of clean Mg(001) surface. {The reported analyzes suggest, that this general trend stems from a strong charge transfer from Mg to Au and consequent enhancement of a surface dipole.} The calculated properties of the AuMg alloy well agree to the experiment. The reported results may readily find application in Au/Mg/AuMg surface physics and technology of metal/semiconductor contacts.   
\end{abstract}
\maketitle

\section{Introduction}
The contacts between metals and semiconductors (M/S) play an important role in many practical electronic devices with far reaching applications. A controlled tuning of contact properties~\cite{Ivanco2000,Xi2014}, that is of technological relevance, relies on the fundamental understanding of the underlying physics and material properties.

Recent progress in planar GaAs Schottky barrier diodes, primarily developed for the X-ray detection applications, stimulated interest in the low work function M/S contacts, since the diodes possessing them revealed unexpectedly and favorably low leakage current~\cite{Bohacek2008,Dubecky2009,Dubecky2013}. Further improvement of these devices with a promising application potential in sensorics, $\gamma$/X-ray detection and medical imaging~\cite{Zatko2004}, relies on the properties of the interfaces between the  reactive low work function metal layers and surface passivisation metals. 
In one of the most striking cases~\cite{Dubecky2013}, the Mg overlayer on GaAs was passivated by the Au. Since the XPS analysis of the formed contact suggests a presence of a non-ideal disordered AuMg alloy~\cite{Dubecky2014} with the changing composition instead of distinct Au/Mg phases, the question remains to be answered: what is the work function of the alloy right at the alloy/GaAs interface? This value would be decisive for further understanding of the non-trivial transport phenomena observed~\cite{Dubecky2013}.

An AuMg alloy is resistant to oxygen, water and organic solvents~\cite{Oyamada2005} and it has been successfully used e.g. as a cathode material in organic light emitting diodes~\cite{Tang1987,Oyamada2005}. To best of our knowledge, its work function is known only from one experiment~\cite{Oyamada2005} and conclusive theoretical work on this subject is not 
available. In order to fill this gap, in the present work we theoretically study \mbox{Au/Mg/AuMg} surface models with the primary goal to understand how the work function of these systems behaves in non-ideal scenarios. To this end, we consider a defective Mg decorated Au(100) surface, Au decorated Mg(001) surface, and various low-index ideal/disordered CsCl-like~\cite{Thomas1986} AuMg surfaces. The selected systems help to clarify how is the work function of the Au-Mg systems affected by the structure, defects and surface composition.

As we show below, within the set of the considered models, the work functions are Mg-dominated. I.e. they always lie below the average of Au and Mg work functions, and mostly close to the Mg one, in agreement with the experimental observations. {We find, that this trend is caused by the surface dipole formation due to the charge transfer from Mg to Au.} The reported results may find practical applications in Au/Mg/AuMg surface physics and in the technology of low-work function M/S contacts.

\section{Methods}

The calculations were performed using the plane-wave density functional (DFT) theory with the ab-initio Perdew-Wang (PW91) exchange correlation (xc) functional~\cite{PW91}, as implemented in the VASP code~\cite{VASP}. {Initial tests of the PBE xc functional~\cite{Perdew1996} are reported in addition.} The nuclei were represented by the projected augmented wave pseudopotentials~\cite{Blochl1994,Kresse1999} and electronic wave functions were expanded in the plane-wave basis set with the 400~eV energy cutoff. The forces acting on nuclei in optimizations were converged to 0.001 eV/\AA. 

The conventional cell parameters of the bulk Au (fcc), Mg (hcp) and AuMg (CsCl structure) were optimized using a 10$\times$10$\times$10 k-point grid and used to construct 1$\times$1 or 2$\times$2 (to keep the even number of electrons per cell) surface five-layer (enough for our purposes~\cite{Fall1999}) symmetric slab models with at least 20\AA~vacuum region. Atomic positions were subsequently optimized using a 10x10x1 k-point grid, whereas the final single-point runs were obtained with the 14x14x1 {(or equivalent)} k-point grids.

The work-functions~($\Phi$) were obtained from the differences between the plane-averaged local potential~{(cf. e.g. Ref.~\citenum{Rusu2006})} energies at the vacuum level ($E_{\textrm{vac}}$) and the Fermi energy ($E_{\textrm{F}}$), i.e. $\Phi=E_\textrm{vac}-E_\textrm{F}$.

{The surface dipole properties are understood in terms of the relative changes of the surface dipole component perpendicular to the ideal optimized surfaces, $\Delta\mu_\bot$, calculated as a dipole correction ~\cite{Makov1995}. A positive $\Delta\mu_\bot$ implies lower $\Phi$ with respect to the reference surface.} 


The AuMg surface energies $\gamma$ (in the limit of $T\rightarrow0$), that determine the stability, were estimated from thermodynamic considerations~\cite{Reuter2001,Reuter2008,Bendavid2013} assuming limiting cases of Au- ($\gamma^{\text{Au-rich}}$) and Mg-rich ($\gamma^{\text{Mg-rich}}$) atmospheres. The Au-rich limit is relevant to the experimental conditions of the GaAs/Mg/Au contact preparation that is of our interest~\cite{Dubecky2013}. We neglect $pV$ terms and zero-point vibrations~\cite{Reuter2001} and the Gibbs free energies are approximated by the total energies ($E$) from the zero-temperature DFT calculations~\cite{Bendavid2013}.

%
The chemical potential of Au (per atom) in the Au-rich limit is solely determined by the Au bulk which is the preferred phase for Au, i.e. 
\begin{equation}
 \mu_{\text{Au}}\approx E_{\text{tot}}^{\text{Au}}/N_{\text{Au}},
\end{equation}
where $E_{\text{tot}}^{\text{Au}}$ is the total Au bulk energy and $N_{\text{Au}}$ is the number of Au atoms per bulk simulation cell.
The chemical potential of Mg is subsequently fixed by the thermodynamic equilibrium condition,
\begin{equation}
 \mu_{\text{Mg}}^{\text{Au-rich}}=\mu_{\text{AuMg}}-\mu_{\text{Au}},
\end{equation}
leading to the approximation
\begin{equation}
 \mu_{\text{Mg}}^{\text{Au-rich}}\approx E_{\text{tot}}^{\text{AuMg}}/N_{\text{AuMg}}-E_{\text{tot}}^{\text{Au}}/N_{\text{Au}}. 
\end{equation}
The surface energy of AuMg expressed in terms of the approximate chemical potentials defined above reads
\begin{equation}
\gamma^{\text{Au-rich}}\approx \frac{1}{2A}[E_{\text{tot}}^{\text{slab}}-N^{\text{slab}}_{\text{Au}}\mu_{\text{Au}}-N^{\text{slab}}_{\text{Mg}}\mu_{\text{Mg}}^{\text{Au-rich}}],
\end{equation}
where $A$ is the slab-model surface area, the factor $1/2$ accounts for the two surfaces in the simulation cell, $N^{\text{slab}}_{\text{Au}}$ and  $N^{\text{slab}}_{\text{Mg}}$ are counts of the number of Au and Mg atoms in the relaxed slab, respectively.

In a similar way, one can define the following chemical potentials relevant for the Mg-rich case,
\begin{equation}
 \mu_{\text{Mg}}\approx E_{\text{tot}}^{\text{Mg}}/N_{\text{Mg}},
\end{equation}
\begin{equation}
 \mu_{\text{Au}}^{\text{Mg-rich}}\approx E_{\text{tot}}^{\text{AuMg}}/N_{\text{AuMg}}-E_{\text{tot}}^{\text{Mg}}/N_{\text{Mg}}, 
\end{equation}
and use them to obtain
\begin{equation}
\gamma^{\text{Mg-rich}}\approx \frac{1}{2A}[E_{\text{tot}}^{\text{slab}}-N^{\text{slab}}_{\text{Au}}\mu_{\text{Au}}^{\text{Mg-rich}}-N^{\text{slab}}_{\text{Mg}}\mu_{\text{Mg}}],
\end{equation}
a surface energy in the limit of Mg-rich atmosphere.

\section{Models}
The studied surface models (cf. Tab.~\ref{tabresA}, Tab.~\ref{tabresB} and Fig.~\ref{fig_disord}) include~\cite{suppInfo}:

i) {clean Au(100) (A1), Au(100) with surface missing-atom defects~(A2-A4), and Au(100) with a subsurface missing atom defect (A5).}

ii) Au(100) surface with one Au atom substituted by Mg (B1), Mg decorated Au(100) with Mg at the top, bridge and hollow positions (B2-B4), B1 with an additional Mg atom on top of Au, i.e. Mg$_2$/Au(100) (B5-6) and Mg(1ML)/Au(100), i.e. Au covered by a single Mg monolayer (B7).   

iii) In the case of Mg, the studied models include a pure Mg(001) surface (M1), Mg(001) with a missing-atom defect (M2).

iv) Concerning the Au decorated Mg, only a single Au atom decorated Au/Mg(001) model (N1) where the Au atom resides at the hollow site is considered. The calculations starting from the top and bridge configurations converged to the same state as N1 and are therefore not reported.

v) The models of AuMg surface include ideal CsCl-like (100), (110), and (111) surfaces (G1-G5), Mg decorated AuMg(110) (G6), and two disordered $3\times3$ (110) surfaces (G7-8) generated by the Born-Oppenheimer {ab-initio} molecular dynamics performed at 2500~K~{\cite{Konopka2009}} and subsequent optimization of the two random snapshots to their respective nearest local minima.  

\section{Results and Discussion}

{\subsection{Benchmarks}}

The bulk lattice parameters of Au and Mg and work functions of clean Au(100) and Mg(001) surfaces, {calculated using PW91 and PBE xc correlation functionals}, are reported in the Tab.~\ref{tabref}. The  results {from both functionals } well agree to the experimental data and previous theoretical calculations~\cite{Prada2008}, confirming the reliability of {the used DFT} approach {(for an extended discussion regarding suitability of DFT functionals for modeling of Au, cf. e.g. Ref.~\citenum{Dubecky2012}).  The data also reveal that the PBE functional performs slightly better with respect to the reported experiment, nevertheless, since both approaches generate similar results and the PW91 is known to perform well in various solid-state surface models containing Au/Mg~\cite{Giordano2005,Rusu2006,Prada2008}, we continue to use PW91 throughout the work.}
\begin{table}[h!]
\scriptsize
\caption{The calculated lattice parameters ($a_0$, $c_0$) of Au (fcc), Mg (hcp), and the work functions ($\Phi$) of Au(100) and Mg(001) surfaces, compared to the experiment.}
\begin{tabular}{llcccccc}
\hline
\hline
           &       &$a_0$/\AA & $c_0$/\AA & $\Phi$/eV \\
\hline
Au/Au(100) & PW91  & 4.18     & -        & 5.10 \\
           & {PBE}   &   {4.16}   &  -    & {5.11} \\
           & experiment  & 4.08$^a$ & -        & 5.22$^b$ \\\\
Mg/Mg(001) & PW91  & 3.20     & 5.18     & 3.72 \\
           & {PBE}   &  {3.20}    &   {5.19}   & {3.72} \\
           & experiment  & 3.21$^c$ & 5.21$^c$ & 3.66$^b$ \\
\hline\hline
$^a$~Ref.~\citenum{Maeland1964}\\
$^b$~Ref.~\citenum{Michaelson1977}\\
$^c$~Ref.~\citenum{Walker1959}\\
\end{tabular}
\label{tabref}
\end{table}

{\subsection{Clean and Au/Mg Decorated Au(100) and Mg(001) Surfaces}}

The top-view illustrations~\cite{VESTA} of the considered Au(100) and Mg(001) surface models together with the calculated and experimental $\Phi$ (where available) {and $\Delta\mu_\bot$} are summarized in the Tab.~\ref{tabresA}. Apparently, the surface defects (missing atoms) lower $\Phi$ with respect to the ideal surface in both, Au (A2-4 vs. A1) and Mg (M2 vs. M1), respectively. The presence of a subsurface defect may, on the other hand, slightly increase $\Phi$ (A5). {These changes  correlate with $\Delta\mu_\bot$ defined with respect to the clean Au (A1), as reported for A2 and A5 and clean Mg (N1), respectively.}

\begin{table}[ht!]
\scriptsize
\caption{The illustrations of the considered Au and Mg surface models, the related work functions ($\Phi$) {and surface dipole differences ($\Delta\mu_\bot$)}. Colors: yellow - Au, blue - Mg belonging to the contiguous surface layer, royal blue - Mg ad-atoms. 
}
\newcolumntype{C}{>{\centering\arraybackslash} m{1.1cm} }
\begin{tabular}{Cm{3.3cm}Cc|C|c}
\hline
\hline
Label&Model & Top view & & $\Phi$/eV & {$\Delta\mu_\bot$/eV\AA{}} \\
\hline
&&&& \\
& Au  {(experiment)}               && & 5.22$^a$  \\
A1& Au(100)                      & \includegraphics[width=25pt,bb=0 0 344 344]{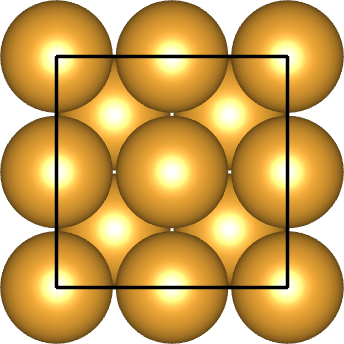}           & & 5.10 \\
A2& Au(100) w/ defect 1          & \includegraphics[width=25pt,bb=0 0 314 314]{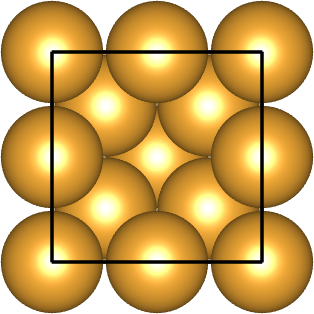}       & & 5.06 & {0.03}\\
A3& Au(100) w/ defect 2          & \includegraphics[width=25pt,bb=0 0 344 344]{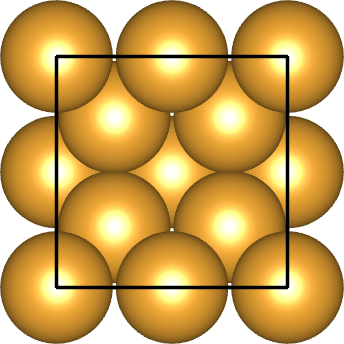}       & & 4.95 \\
A4& Au(100) w/ defect 3          & \includegraphics[width=25pt,bb=0 0 314 314]{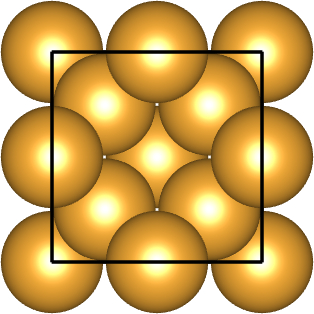}       & & 4.75 \\
A5& Au(100) w/ bulk defect       & \includegraphics[width=25pt,bb=0 0 344 344]{./Au100.jpg}           & & 5.21 & {-0.02}\\
\hline
&&&&\\
B1& Au(100) Mg substituted        & \includegraphics[width=25pt,bb=0 0 304 304]{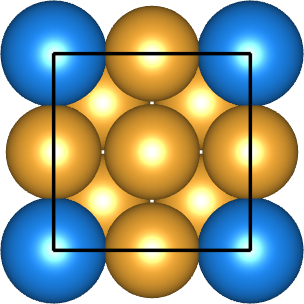} & & 4.59 & {0.14}\\
B2& Au(100)/Mg top                & \includegraphics[width=25pt,bb=0 0 356 356]{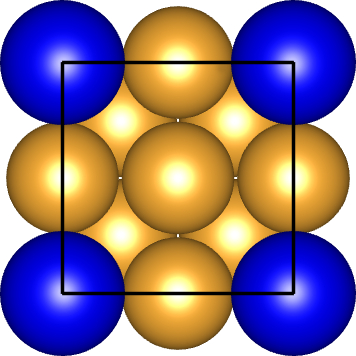}     & & 3.57 & {0.47}\\
B3& Au(100)/Mg bridge             & \includegraphics[width=25pt,bb=0 0 311 320]{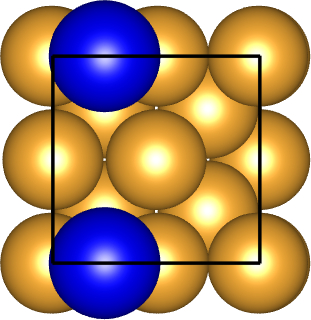}   & & 3.46 & {0.50}\\
B4& Au(100)/Mg hollow             & \includegraphics[width=25pt,bb=0 0 311 311]{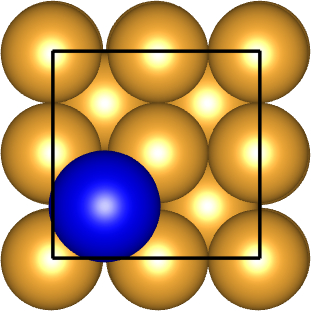}   & & 3.40 & {0.48}\\
B5& Au(100)/Mg$_2$ X              & \includegraphics[width=25pt,bb=0 0 392 392]{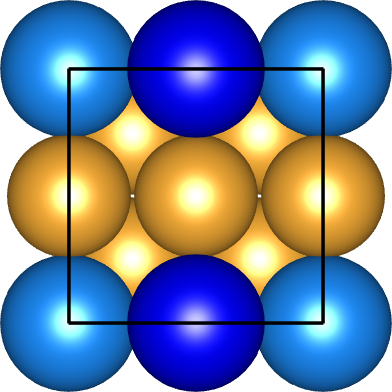}   & & 3.42 & {0.48}\\
B6& Au(100)/Mg$_2$ XY             & \includegraphics[width=25pt,bb=0 0 392 392]{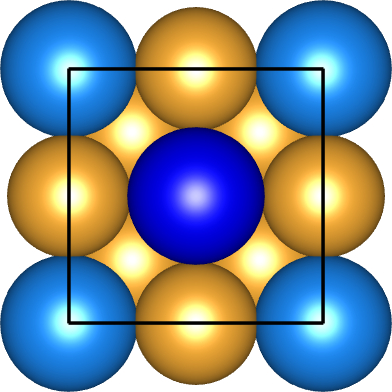}   & & 3.37 & {0.48}\\
B7& Au(100)/Mg(1ML)               & \includegraphics[width=25pt,bb=0 0 344 344]{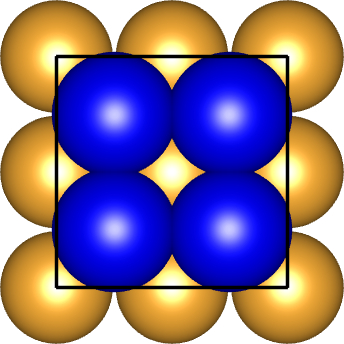}      & & 3.87 & {0.38}\\
\hline
&&&&\\
   & Mg  {(experiment)}                                                                     & & & 3.66$^a$\\
M1 & Mg(001)                      & \includegraphics[width=34pt,bb=0 0 794 541]{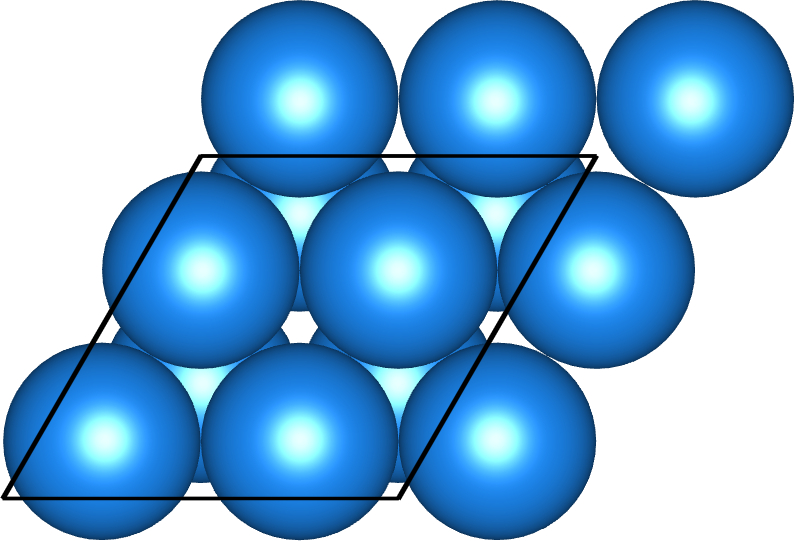}          & & 3.72 \\
M2 & Mg(001) w/ defect            & \includegraphics[width=34pt,bb=0 0 936 636]{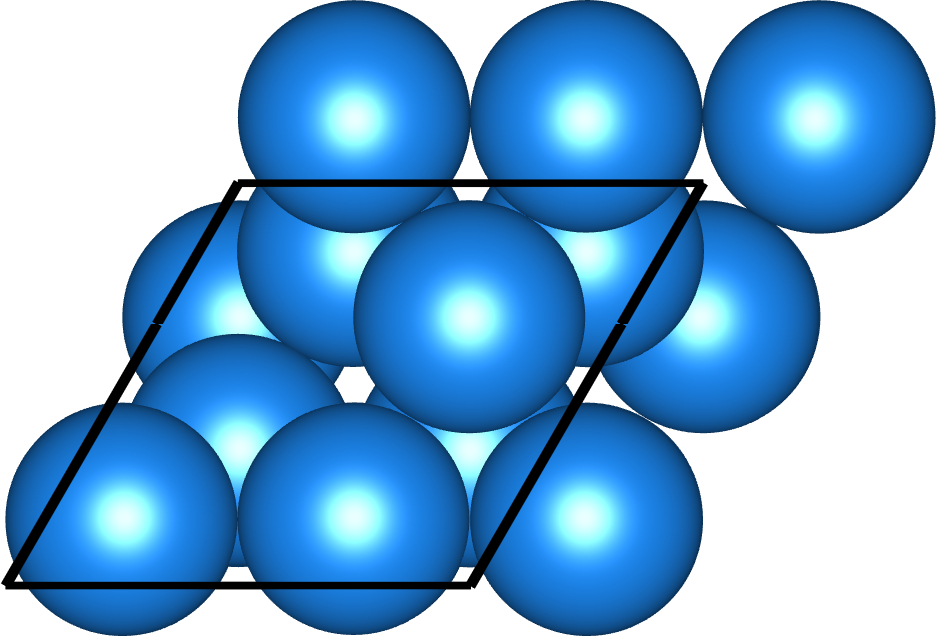}      & & 3.71 \\
\hline
&&&&\\
N1 & Mg(001)/Au hollow            & \includegraphics[width=34pt,bb=0 0 786 535]{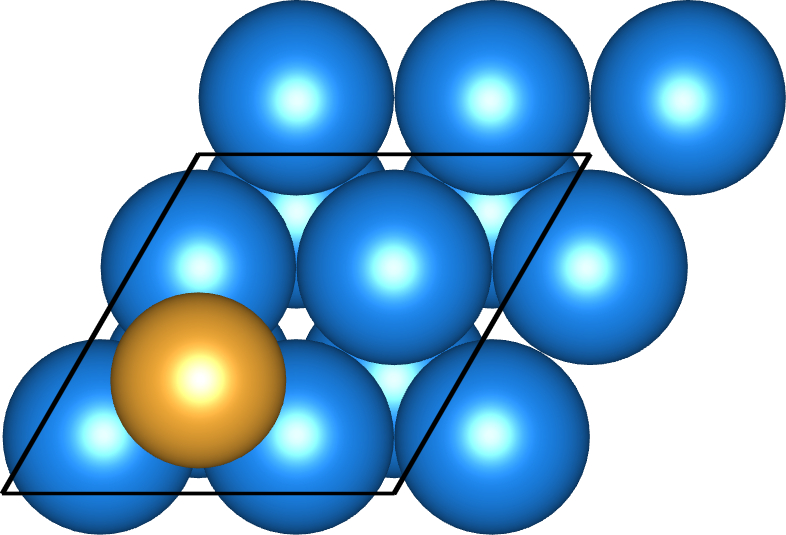}& & 4.10 & {-0.12}\\
\hline
\hline
$^a$~Ref.~\citenum{Michaelson1977}\\
\end{tabular}
\label{tabresA}
\end{table}

In the case of Mg-substituted Au(100) surface (B1), where one of the surface Au atoms per cell is replaced by Mg, the surface $\Phi=4.59$~eV is significantly lowered (by 0.51~eV) with respect to the clean Au(100) ($\Phi=5.10$~eV, A1). The work functions drop further {by an additional $\sim$~1~eV,} to $\Phi=3.37-3.57$~eV, when Mg decorates the Au surface. In all the considered cases {where Mg decorates Au} (B2-B6), these values are even below the {theoretical value of} $\Phi$ of an ideal clean Mg(001) surface ($\Phi=3.72$~eV).

The work function of the Au surface covered by an ideal single Mg monolayer, i.e.  Mg(1ML)/Au(100) (B7), amounts to 3.87~eV. {Here, the drop of $\Phi$ is not as pronounced, since $\Delta\mu_\bot$ induced by the adsorbed {\it ideal} monolayer of Mg is smaller than in the cases where Mg atom lies on the clean/substituted Au surface (B2-B6)}. 

For completeness, we mention that the Au decorated Mg(001) reveals the $\Phi=4.10$~eV, i.e. higher by 0.44~eV with respect to the clean Mg, nevertheless still below the average of clean Au and Mg~($\Phi_{\text{avg}}$=4.41~eV). The data clearly indicate, that Mg strongly dominates the surface work functions of Au/Mg surfaces containing Mg. 

\begin{figure}\label{figDOS1}
 \centering
 \includegraphics[width=240pt, bb= 0 0 1600 798]{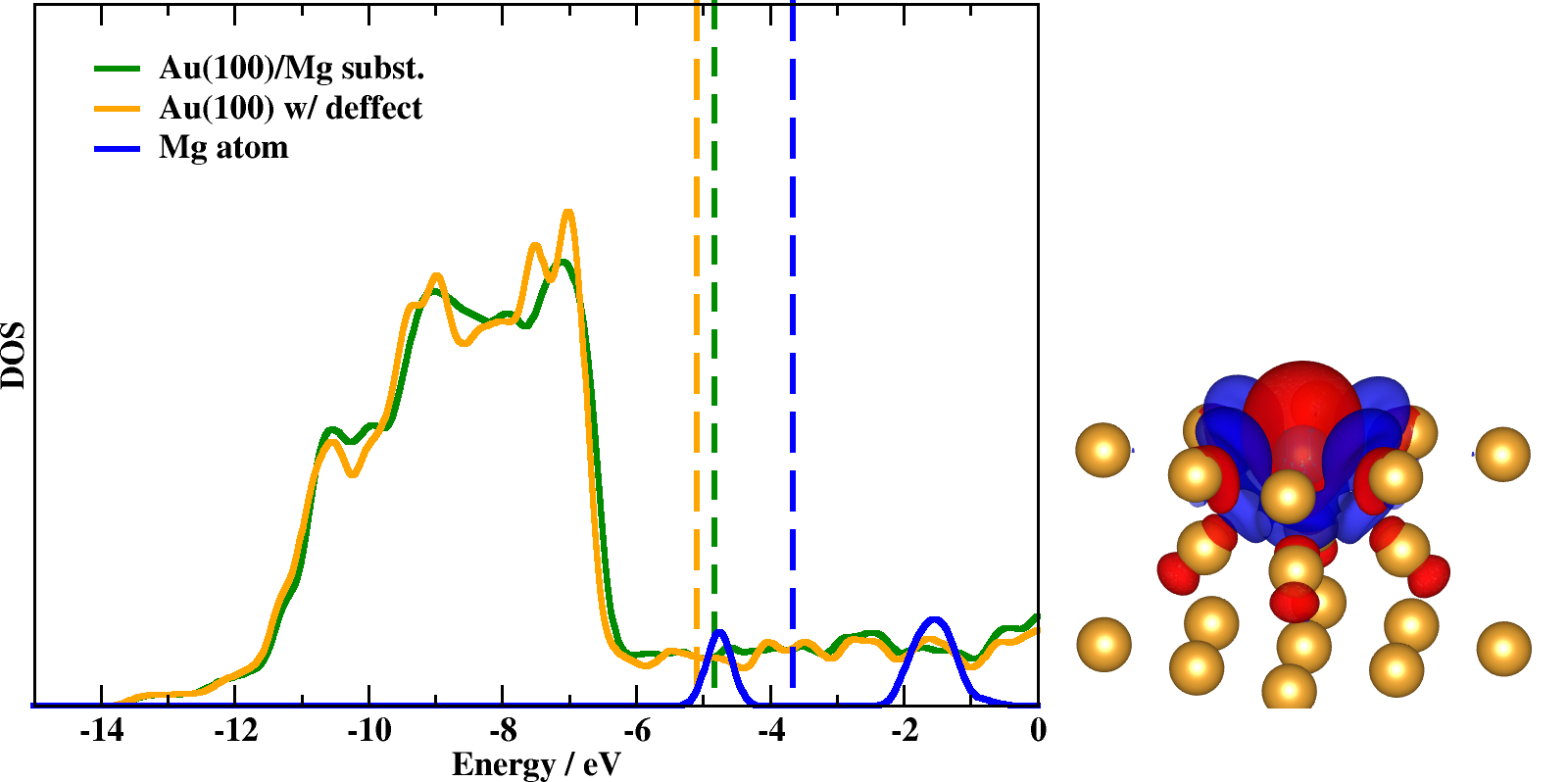}
 \newline
 \newline\\
 \includegraphics[width=240pt, bb= 0 0 1600 798]{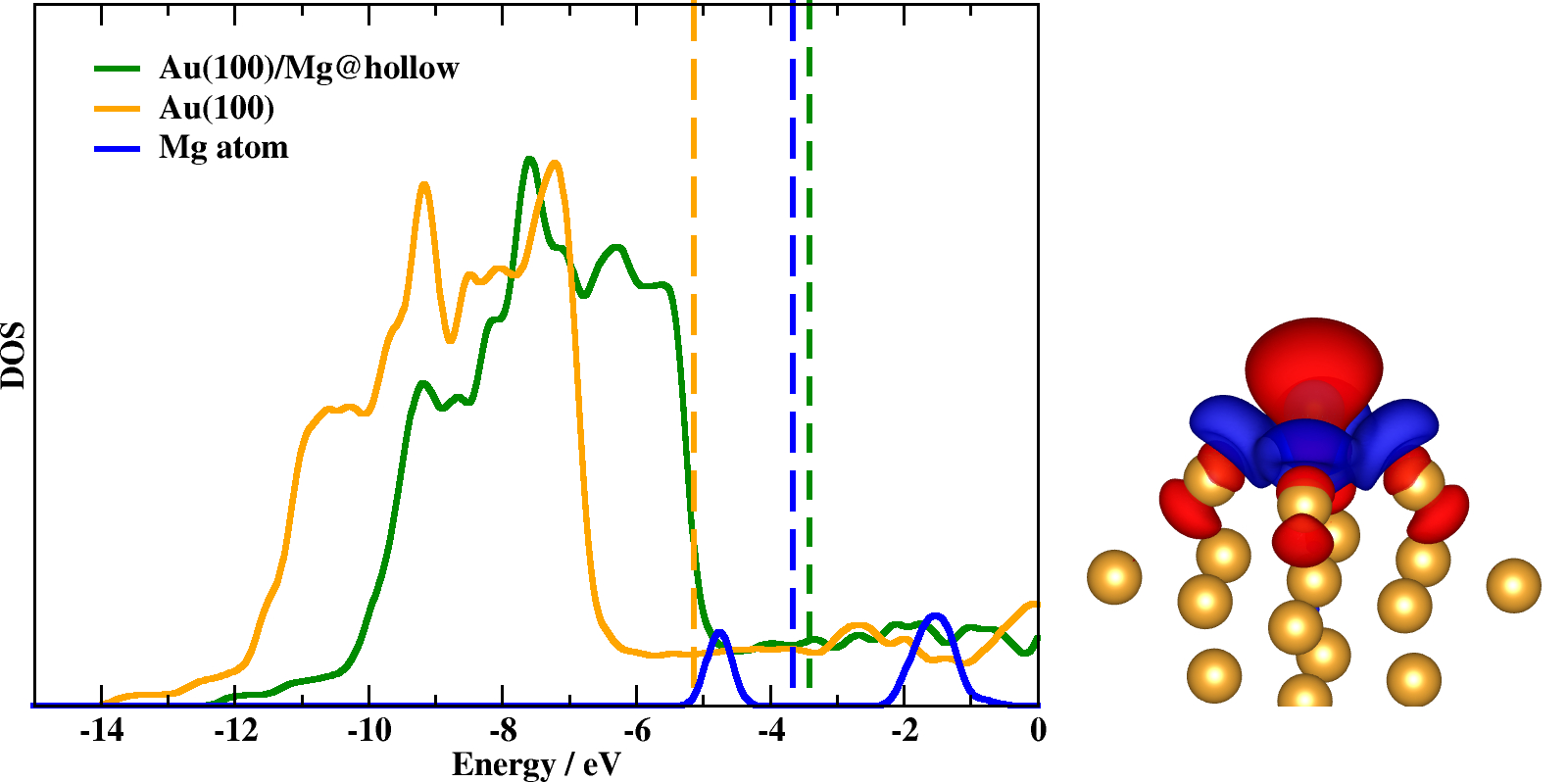} 
 {\caption{Densities of states (DOS) for the B1 (top) and B4 (bottom) models of the Mg-modified Au surface (cf.~Tab.~\ref{tabresA}). The vacuum energy is aligned to zero and the vertical lines indicate the Fermi level. The charge density differences (right; decrease - red, increase - blue, isosurface 0.0015 a.u.) were obtained from the final Mg-modified Au(100) surface models and their pure unoptimized Au/Mg constituents in the identic cell.}}
\end{figure}

{In order to understand why the presence of Mg induces such a strong lowering of the Au surface work function, we further analyze the selected representative models, B1 and B4, in terms of the density of states~\cite{Giordano2005} (DOS) and electron density differences (Fig.~1). The Fermi levels in Fig.~1 in both cases, B1 and B4, indicate that the charge must flow from Mg to Au in order to maintain equilibrium. This effect is demonstrated in the electron density difference plots (Fig.~1), which reveal a qualitative difference between the models, even though the $s$ charge transfer from Mg due to the reaction with Au is very similar (0.18$e$ in B1 vs. 0.13$e$ in B4; estimates from incomplete projections within the Wigner-Seitz radii) and even slightly more pronounced in case with higher $\Phi$. In B1 where one of the surface atoms is substituted by Mg, the charge density is primarily rearranged within the surface plane, whereas in B4, the rearrangement takes place primarily along the surface 
normal, which leads to a more enhanced $\Delta\mu_\bot$, thus lowering $\Phi$ more significantly. } 

\begin{figure} \label{figcorr}
 \centering
 \includegraphics[width=175pt,bb=0  0 1150 793]{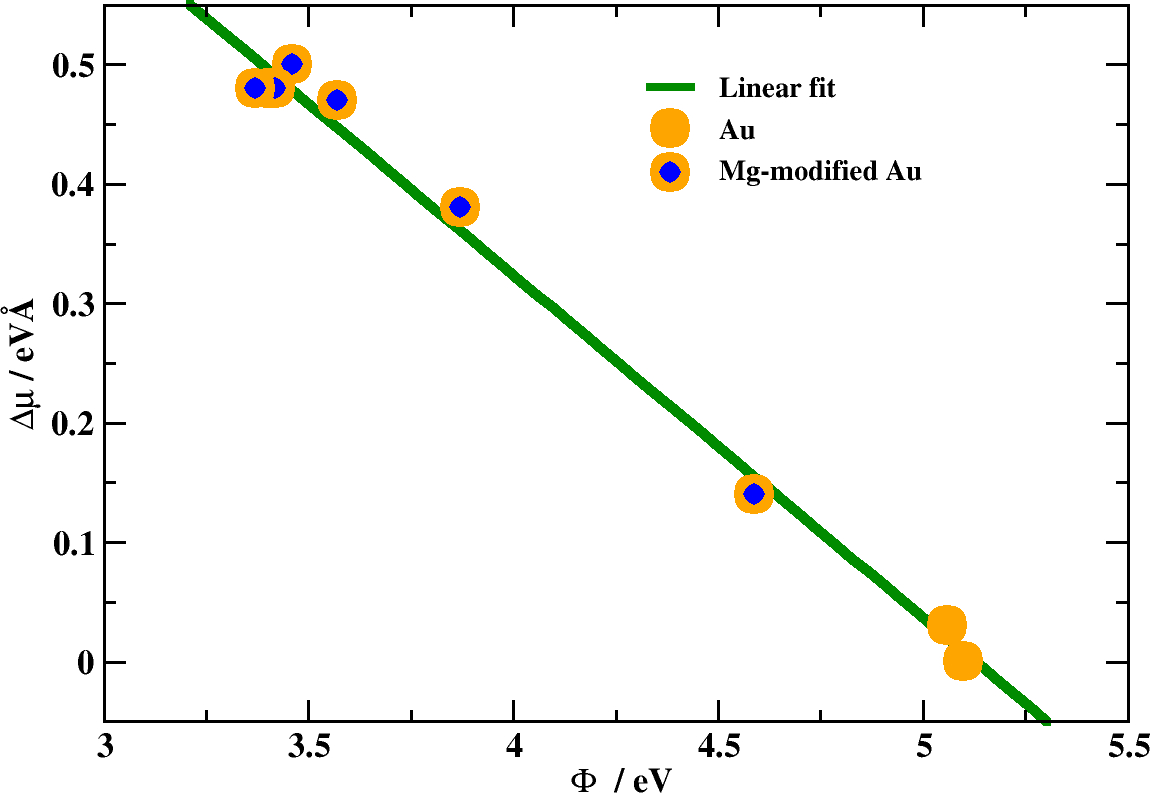}
{ \caption{Dependence of the work functions $\Phi$ on the relative out-of-surface dipole component $\Delta\mu_\bot$ in the studied Mg-modified Au surfaces.}}

\end{figure}

{An effect of the surface dipole and its directionality is thus identified as a primary reason responsible for the significant lowering of $\Phi$ observed in all the Mg-modified Au surface models (B1-B7), which is further corroborated by the remarkably good correlation between $\Delta\mu_\bot$ and $\Phi$, as reported in Fig.~2.}


{\subsection{AuMg Alloy Surfaces}}

{The optimum lattice constant of bulk AuMg with CsCl structure was found to be $a=3.31$~\AA{}, in agreement with the previous theoretical and experimental values~\cite{Methfessel2000}.}

The results concerning the studied low-index CsCl-like AuMg surfaces,  together with the related illustrations, are reported in the Tab.~\ref{tabresB}. The Mg-rich (G1) and Mg-top (G4) surfaces reveal the lowest $\Phi$, and the Au-rich and Au-top surfaces (G2 and G5) reveal the highest $\Phi$, as expected. The surface energies $\gamma${, that determine the surface stability at realistic conditions,}  in both considered limits (Au-rich and Mg-rich {atmospheres}) reveal that the most stable surface is AuMg(110), containing Au and Mg in the same plane (G3, $\Phi=4.12$~eV).

{In the following, we further analyze the most thermodynamically stable AuMg(110) surface. First, we consider an additional Mg adatom on the G3 surface, that leads to the G6 model. The Mg adatom further lowers $\Phi$ by a significant amount of 0.53~eV, that is fully attributable to the change in the surface dipole component $\Delta\mu_\bot$ (Tab.~\ref{tabresB}).}

\begin{figure}[t!]
 \centering
 \includegraphics[width=98pt,bb=0 0 1000 500]{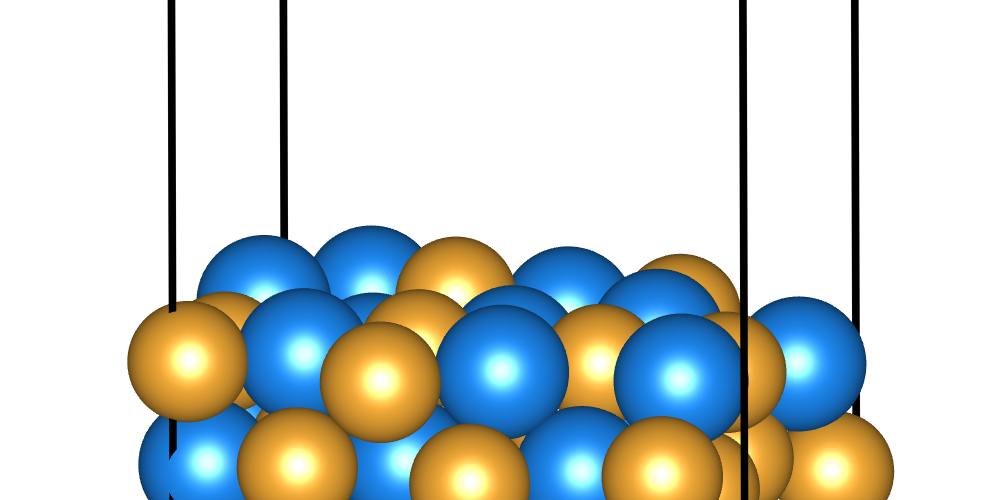} (G7)
 \includegraphics[width=98pt, bb= 0 0 1000 500]{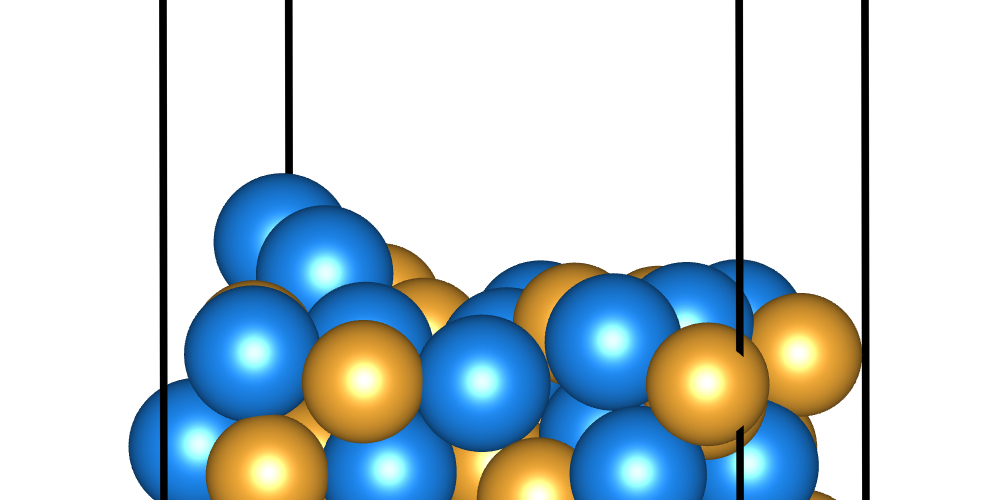} (G8)
 \caption{The disordered AuMg(110) 3$\times$3  surface models produced by molecular dynamics (at 2500~K) and subsequent optimization to the nearest local minimum on the potential energy surface.}
 \label{fig_disord}
\end{figure}

{Finally, an effect of disorder on the work function of AuMg(110) is considered. In order to produce non-ideal} structures, the AuMg(110) was annealed at \mbox{2500~K} and two randomly chosen snapshots were subsequently reoptimized (G7 and G8, Fig.~\ref{fig_disord}). A resulting disorder{ed structures reveal} lowering {of} $\Phi$ from the original 4.12~eV to 4.04 (G7) and 3.82~eV (G8), respectively.  
In the case of G8, a more pronounced lowering of $\Phi$ is attributed to the presence of Mg dimer lying out-of-plane, a structural feature of this model { (qualitatively similar to the B5 and B6 models of Mg-modified Au)}. Therefore, a nonideality enables an easier electron withdrawal (lower $\Phi$) from the surface, similar to the Mg adatom cases (B3, B4-6 and G6), compared to the the ideal case (G3) or non-ideal case (G7) that is {more} flat {(or closer to the ideal clean surface)}.

\begin{table*}[ht!]
\scriptsize
\caption{The illustrations of the considered AuMg surface models, the related work functions ($\Phi$), surface energies, $\gamma^{\text{Au-rich}}$ and $\gamma^{\text{Mg-rich}}$ {and surface normal dipole component changes $\Delta\mu_\bot$} (for definitions, cf. Methods). Colors: yellow - Au, blue - Mg.
}
\newcolumntype{C}{>{\centering\arraybackslash} m{1.4cm} }
\newcolumntype{D}{>{\centering\arraybackslash} m{2.05cm} }
\begin{tabular}{Cm{3.3cm}Cc|C|Dc|c}
\hline
\hline
Label&Model & Top view & & $\Phi$/eV & $\gamma^{\text{Au-rich}}$/eV\AA{}$^{-2}$ & $\gamma^{\text{Mg-rich}}$/eV\AA{}$^{-2}$ & {$\Delta\mu_\bot$/eV\AA{}} \\
\hline
&&&&&&\\
  & AuMg {(experiment)}         &                        & & 3.70$^a$ & & \\
G1& AuMg(100) CsCl, Mg-rich   & \includegraphics[width=27pt,bb=0 0 506 506]{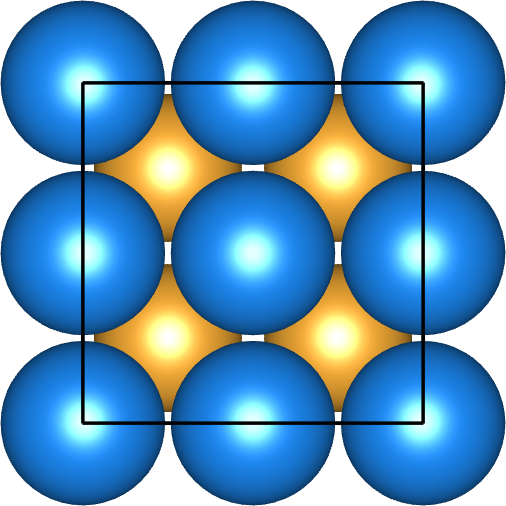}  & & 3.37 & 1.23  & 0.08 \\
G2& AuMg(100) CsCl, Au-rich   & \includegraphics[width=27pt,bb=0 0 538 538]{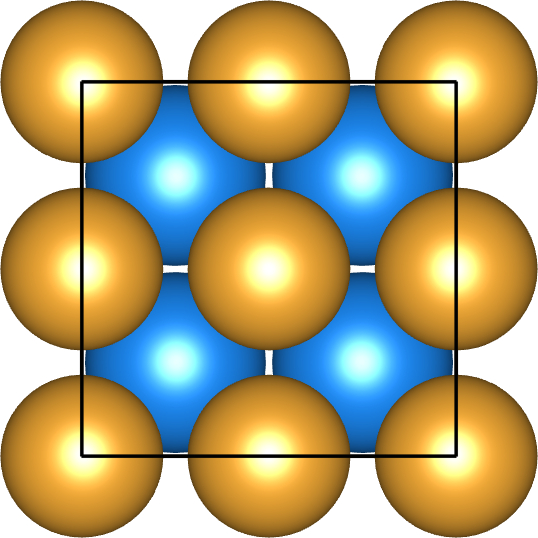}  & & 4.83 & 1.08  & 0.15 \\
G3& AuMg(110) CsCl            & \includegraphics[width=23pt,bb=0 0 346 272]{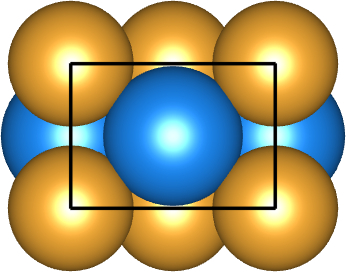}    & & 4.12 & 0.73  & 0.06 & \\
G4& AuMg(111) CsCl, Mg-top    & \includegraphics[width=30pt,bb=0 0 591 419]{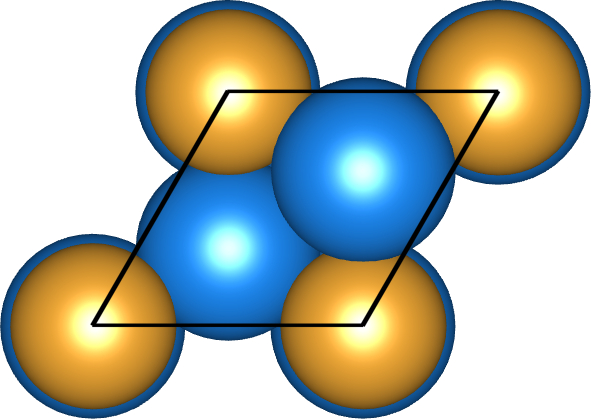}   & & 3.58 & 1.02$^b$ & 0.09$^b$\\
G5& AuMg(111) CsCl, Au-top    & \includegraphics[width=30pt,bb=0 0 651 462]{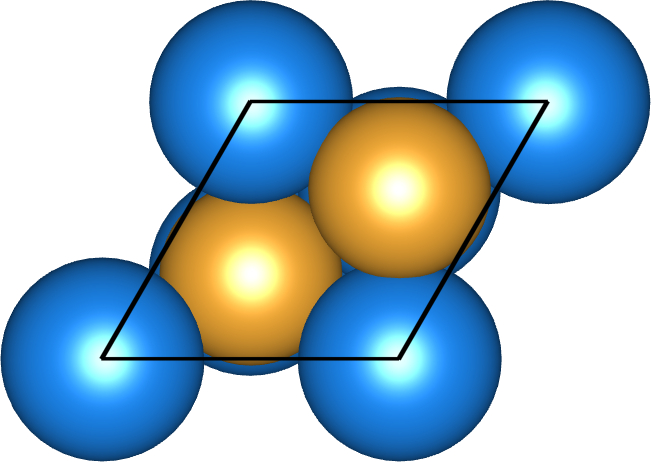}   & & 4.28 & 1.02$^b$ & 0.09$^b$\\
&&&&&&\\
\hline&&&&&&\\
G6& AuMg(110) Mg ad-atom & \includegraphics[width=22pt,bb=0 0 792 544]{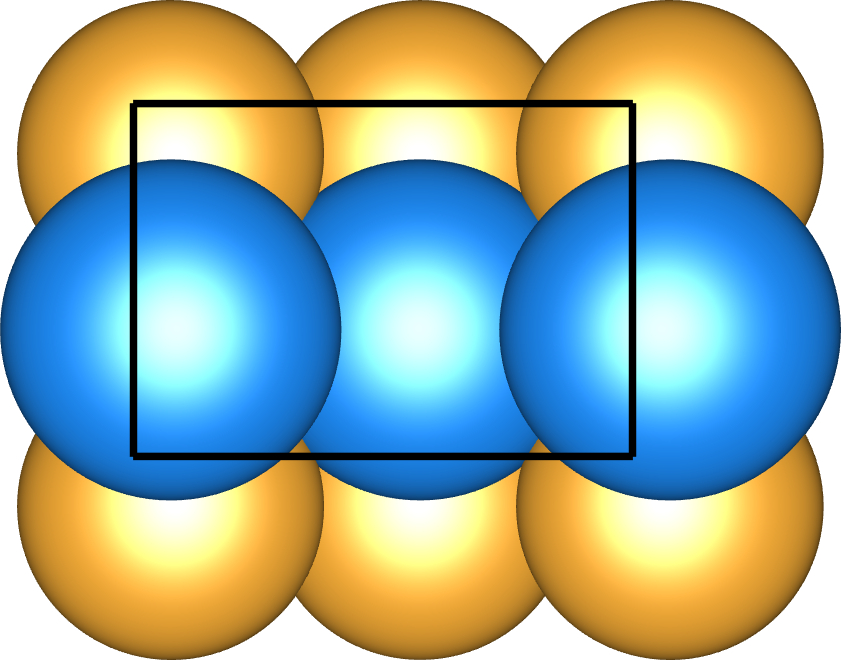} & & 3.59 & -  &  - & {0.10}\\
&&&&&&\\
G7& AuMg(110) disordered      & Fig.~\ref{fig_disord}         & & 4.04 & - & - \\
&&&&&&\\
G8& AuMg(110) disordered     & Fig.~\ref{fig_disord}       & & 3.82 & - &  -\\
&&&&&&\\
\hline\hline
$^a$~Ref.~\citenum{Oyamada2005}\\
$^b$~Average\\
\end{tabular}
\label{tabresB}
\end{table*}

Overall, {the considered AuMg surfaces} (except for the ideal pure Au-terminated surface) show a trend observed in Mg decorated Au, i.e. that the work functions are Mg dominated (lie below $\Phi_{\text{avg}}$) and typically approach {the work function of} pure Mg. Nonideality/disorder further {lowers} $\Phi$. Based on the reported results, {we theorize} that the work function of an amorphous AuMg is similar to the pure Mg, a conclusion in agreement with the experimental observations~\cite{Oyamada2005} and expectations~\cite{Dubecky2013}.


\section{Conclusions}

The work functions of the non-ideal Au, Mg and AuMg surfaces were calculated by the ab-initio plane-wave density functional theory. Irrespective of the details, the considered models containing Mg on the surface, {including} AuMg alloys, reveal strongly Mg-dominated work functions, i.e. significantly shifted toward the work function of pure Mg(001) surface. {This effect is dominantly caused by the enhancement of the surface dipole due to the charge transfer from Mg to Au and more pronounced if the charge transfer occurs along the surface normal direction.}  A stable AuMg alloy possesses a low work function similar to the reactive Mg, while being remarkably stable against water and air~\cite{Oyamada2005}, and is thus well suited for technological applications including contact metallizations in devices based on metal/semiconductor interfaces.

\begin{acknowledgments}
The support from the Operational Programme Research and Development (OP R\&D) for Innovations - European Regional Development
Fund (ERDF, project CZ.1.05/2.1.00/03.0058), the OP Education for
Competitiveness - European Social Fund (projects CZ.1.07/2.3.00/30.0004 and CZ.1.07/2.3.00/20.0058), and by the Slovak Grant Agency for Science (VEGA 2/0167/13 and 2/0175/13), is gratefully acknowledged.
The calculations were in part performed at the Slovak HPC infrastructure (projects ITMS 26230120002 and 26210120002)
supported by the ERDF OP R\&D.

\end{acknowledgments}


\begin{thebibliography}{10}
\bibitem{suppInfo}
The optimized structures are available~upon request.
\end{thebibliography}

\end{document}